# A comparison of stochastic and deterministic downscaling in eddy resolving ocean modelling: the Lakshadweep Sea case study

Georgy I.Shapiro[1], Jose M.Gonzalez-Ondina[2], Mohammed Salim[2], J.Tu[2]

[1] University of Plymouth, Drake Circus, Plymouth PL4 8AA, UK

[2] University of Plymouth Enterprise Ltd, New Cooperage Building, Royal William Yard, Plymouth, UK, PL4 8AA P

Key points:

- The paper compares two non-data assimilating regional models nested into the same data assimilating global model.
- The first model (LD20_SDD) uses a stochastic-deterministic downscaling method while the second model (LD20_NEMO) uses a dynamical downscaling.
- Both LD20 models show similar skills in reproducing temperature and salinity assessed against observations in terms of root-mean-square of anomalies
- LD20_SDD has a better bias
- LD20_SDD is much faster and uses significantly less computational resources than LD20_NEMO.

Corresponding author: Georgy I.Shapiro (g.shapiro@plymouth.ac.uk)

# Abstract
This study compares the skills of two numerical models having the same horizontal resolution but based on different principles in representing meso- and submesoscale features of ocean dynamics in the Lakshadweep Sea (North Indian Ocean). The first model, titled LD20-NEMO, is based on solving primitive equations using the NEMO (Nucleus for European Modelling of the Ocean) modelling engine. The second one, titled LD20-SDD, uses a newer Stochastic-Deterministic Downscaling method. Both models have $1/20^o$ resolution and use the outputs from a Global Ocean Physics Analysis and Forecast model at $1/12^o$ resolution available from Copernicus Marine Service (CMEMS). The LD20-NEMO uses only a 2D set of data from CMEMS as lateral boundary conditions. The LD20-SDD consumes the full 3D set of data from CMEMS and exploits the stochastic properties of these data to generate the downscaled field variables at higher resolution than the parent model. The skills of the three models, CMEMS, LD20-NEMO and LD20-SDD are assessed against remotely sensed and in-situ observations for the four-year period 2015-2018. All models show similar skills in reproducing temperature and salinity, however the SDD version performs slightly better than the NEMO version. This difference in resolution is particularly significant in simulation of vorticity and computation of the share of the sea occupied by highly non-linear processes. While the NEMO and SDD model show similar skill, the SDD model is more computationally efficient than the NEMO model by a large margin.

# Introduction
There is a growing tendency to move to higher and higher resolution in ocean modelling. Higher resolution models are particularly helpful in simulations of ocean circulation in coastal and shelf seas and in the vicinity of intensive jet currents such as the Gulf Stream or Kuroshio (Volkov et al. 2015; Kang and Curchitser 2013; Kerry et al. 2016). The enhanced ability of a model to resolve mesoscale and submesoscale eddies leads to significant improvement in simulation of large scale features such as the Gulf Stream recirculation (Chassignet and Xu 2017). High resolution physical models provide a solid background for the study and prediction of ecosystem dynamics and the distribution and productivity of key marine species with remarkable detail and realism. Such ocean models underpin sustainable resource management, improvement in food security and development of Blue Economies (Solstice 2021). However, higher resolution comes at a cost. It is commonly accepted that the increase of horizontal resolution in ocean models is associated with significant increase in required computing power, typically by a factor of ten for each increase of the horizontal resolution by a factor of two (Chassignet and Xu 2021). The enhancement of resolution by a factor of three from ORCA025 (1/4°) to ORCA12 (1/12°) grid in a global ocean model resulted in the 24-fold increase in computational time on the UK Met Office supercomputer (Hewitt et al. 2021).

Therefore, a development of new time saving algorithms could provide a cost-effective solution in high-resolution modelling. One of such algorithms titled Stochastic-Deterministic Downscaling (SDD) was proposed in (Shapiro et al. 2021). It is based on the philosophy that at smaller scales the ocean processes become more chaotic and resemble to some extent the dynamics of small-scale turbulence which is studied by methods of statistical fluid dynamics (Monin and Yaglom 2013). Hence, there is an intention of simulating small-scale ocean processes employing their stochastic properties inferred from data in addition to deterministic properties inferred from equations of motion. As a source of data the SDD method uses outputs from a coarser resolution (parent) ocean model. In this study we contrast and compare the efficiency and accuracy of the SDD method against traditional deterministic ocean circulation model in the Lakshadweep Sea located in tropical Indian Ocean.

# Materials and methods

The area of study is located within 68E to 78E and 7.50N to 14.50N to the west of Indian peninsula around the Lakshadweep archipelago containing 36 islands, atolls and coral reefs. The parent model is the operational global model at 1/12 degree of resolution and 50 vertical layers available from Copernicus Marine Service, product GLOBAL_REANALYSIS_PHY_001_030 (CMEMS, 2020). This product is not available from CMEMS anymore and has been upgraded to product GLOBAL_MULTIYEAR_PHY_001_030. The parent model assimilates observational data on Sea Surface Temperature (SST), Sea Surface Height, and in-situ Temperature/Salinity profiles. The parent model provides outputs, amongst others, of potential temperature, salinity, meridional and zonal components of velocity. The model outputs are compared to three observational data sets: OSTIA (2022), Argo float temperature/salinity profiles (Argo 2022) and GHRSST Multiscale Ultrahigh Resolution (MUR) L4 analysis (GHR-MUR 2022).

### LD20-SDD model

The child SDD-LD20 model has the same geographical limits as the parent model, however different depth levels, also 50 in number, were selected to be better suited to the dynamics of the Lakshadweep Sea than the parent model. The daily averaged outputs of temperature, salinity and horizontal velocity are obtained by Statistical-Deterministic Downscaling from the parent to the child model. The SDD method (Shapiro et al 2021) is based on the modified version of Objective Analysis (Gandin 1965; Kalnay 2003) which is applied to the parent model output in order to downscale it to a finer (child) model grid. The method treats fluctuations of field variables around their statistical means as a random process to which Markov-Gauss algorithm can be applied in order to minimise, in statistical sense, the error of calculation of field variables on the fine grid. The SDD method assumes isotropy and local spatial homogeneity of the first and second statistical moments of the probability distribution function. Local spatial homogeneity is defined in this case as small relative variations of statistical moments over the length of one grid cell. The method allows to reveal details of oceanic features which are only embryonically represented by the parent model. The SDD method requires knowledge of the correlation functions of fluctuations of field variables. It uses the usual ergodic hypothesis that replaces ensemble averages with time averages (Moore 2015). The slowly changing averages are calculated using a moving time window. The length of the window is chosen to be long enough to have sufficient number of members for averaging but short enough so that the seasonal variability can be ignored.

In this study we used 11 days as the length of the time window and the total time period was two years (01-01-2016 to 31/12/2017). The correlation function is calculated for each field variable $Q$ and for each parent grid node. First, the time averages $\mathrm{E}_w(Q_{ni}^k)$ within the time window centred at time $t^k$ are computed for all the nodes $ni = 1,2 \ldots N_p$, where $N_p$ is the total number of nodes in the parent mesh. The subscript $w$ indicates that time averaging is done only within the temporal window. Then fluctuations $Q_{ni}^{\prime k} = Q_{ni}^k \quad \mathrm{E}_w(Q_{ni}^k)$ are calculated at all grid nodes for the time point $t^k$. Fluctuations related to the same time point $t^k$ but different nodes are used to calculate the products of fluctuations $Q_{ni}^{\prime k} Q_{n0}^{\prime k}$, where $n0$ is the node under consideration or 'central' node. The process is repeated for different 'central' nodes in the 3D parent model domain. Second, the time point $t_k$ (and the related moving average time window) are shifted by one time point (in our case one day) to calculate the next set of averages, fluctuations, and their products. Third, the spatial correlations $\mathrm{Cor}(Q_{n0}^{\prime}, Q_{ni}^{\prime})$ are computed between each 'central' node $n0$ and other grid nodes $ni$ at the same depth level using the equation

$$\text{Cor}(Q'_{n0}, Q'_{ni}) = \frac{\text{E}_t(Q'_{n0} Q'_{ni})}{\text{std}_t(Q'_{n0}) \text{std}_t(Q'_{ni})} \quad (1)$$

where $\text{E}_t$ and $\text{std}_t$ denote averaging and standard deviation respectively which are calculated over a large time period (in our case two years). The process was repeated for different 'central' nodes.

The field variables are correlated through a number of process having different length scales. Following the approach suggested in (Mirouze et al. 2016) we introduce two correlation length scales — $L_s$ and $L_l$ for short-range and a long-range correlations respectively. They are estimated by fitting, at every node, an isotropic Gaussian curve of parameters $a \in [0,1]$ $L_s, L_l > 0$ to the correlation values obtained by Eq. (1).

$$C'(r, \boldsymbol{r_0}) = a(\boldsymbol{r_0}) \exp\left[-\left(\frac{r}{L_s(\boldsymbol{r_0})}\right)^2\right] + (1 - a(\boldsymbol{r_0})) \exp\left[-\left(\frac{r}{L_l(\boldsymbol{r_0})}\right)^2\right] \quad (2)$$

where $\boldsymbol{r_0}$ is the vector of coordinates of the node $n0$, and $r$ is the distance between the nodes $n0$ and $ni$. For eddy-resolving modelling we are interested in the short-range correlation represented by the correlation length $L_s(\boldsymbol{r_0})$, therefore in calculation of correlations using Eq. (1) we only include the nodes $ni$ which belong to the 'search area' around each 'central' node, in this case it was $1.7 \times 1.7$ degrees in size (4–5 times greater than the anticipated short length scale). Once the correlations are computed, only the short-length component of the correlation given in Eq. (3) are used for the downscaling

$$C(r, \boldsymbol{r_0}) = \exp\left[-\left(\frac{r}{L_s(\boldsymbol{r_0})}\right)^2\right], \quad (3)$$

The computations according to Eq. (1) are carried out for a 3D array of 'central' nodes on the parent grid to create a 3D array of correlation lengths. To give a feeling of the numbers, the total number of $n0$ nodes of the parent model within the LD20_SDD domain is 306,106. As expected, the values of $L_s(\boldsymbol{r_0})$ depend only weakly on $\boldsymbol{r_0}$ supporting the assumption of local statistical homogeneity. Therefore, if $\boldsymbol{r_0}$ is the vector of coordinates of a node on the child rather than parent model grid, then the correlation length $L_s(\boldsymbol{r_0})$ can be approximated by its value at neighbouring points. With this in mind, and to reduce the computation times and the effect of outliers, we compute the fitting using Eq. (2) for only every other node in each horizontal dimension of the parent mesh, while the correlation lengths for other nodes are obtained by linear interpolation. The correlations thus computed are smoothed layer-by-layer with a 2D Gaussian filter. This filtering respects the assumption of local statistical homogeneity as the correlation lengths vary smoothly in space (Weaver and Mirouze, 2013).

Figure 1a and Figure 1b show examples of correlation data sets $\text{Cor}(Q'_{no}, Q'_{ni})$ for SST calculated using Eq. (1) at two different locations of the 'central' node $n0$ and the fitted Gaussian curves. For comparison, Gaussian curves corresponding to the short length scales are superimposed. The scattering of correlation coefficients is relatively small within the short (mesoscale) length and become larger at greater distances. Similar graphs (not shown) were obtained for other field variable and other depths levels. The smaller scatter at shorter distances is consistent with greater coherency of ocean structures within meso and sub-mesoscale ranges.

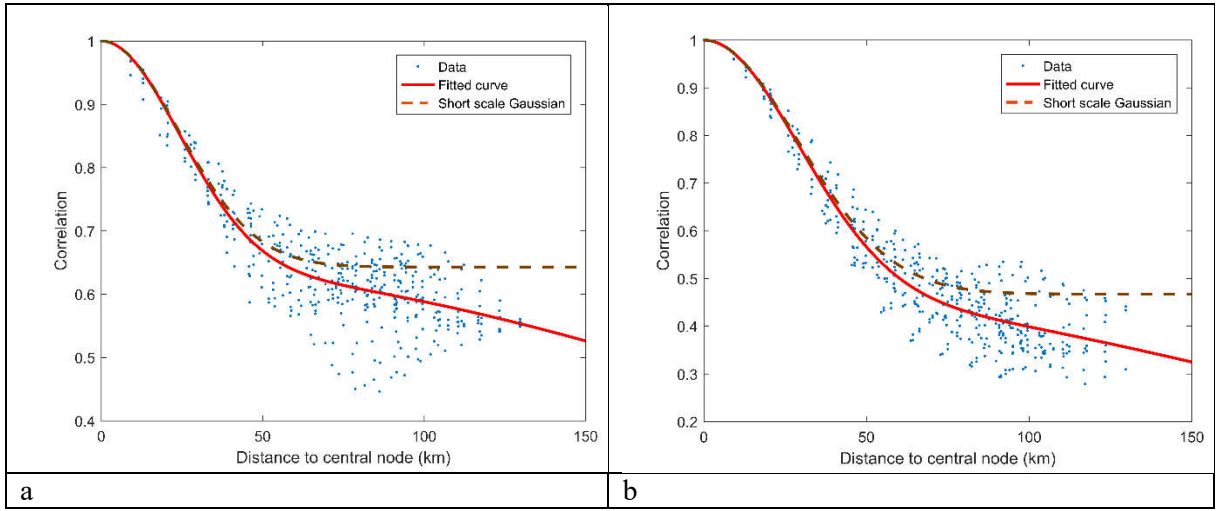

**Fig. 1** Correlation data sets for SST at two locations of the 'central' node n0: (a) 10.7°N, 70.3°E and (b) 7.3°N and 76.2°E. Blue dots represent correlation coefficients of SST fluctuations between the central node and surrounding nodes $ni$. The solid red line represents the fitted two-scale correlation function according to Eq. (2). The dashed line shows a superimposed Gaussian curve corresponding to the short correlation scale in such a way that it can be visually compared to the two-scale one

According to Eq. (2), the correlation function is in general different for different 'central' node $n0$. Figures 2(a-h) show spatial distribution of the correlation lengths across the domain at the surface and at a depth of 156 m for temperature (T), salinity (S), U- and V- component of current velocity. Similar maps are obtained for other depth levels of the parent mesh.

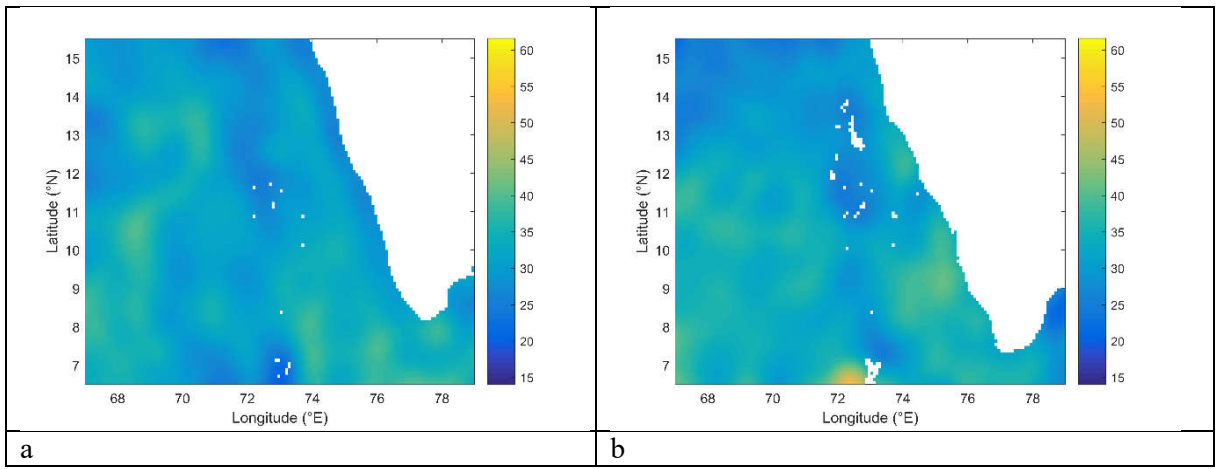

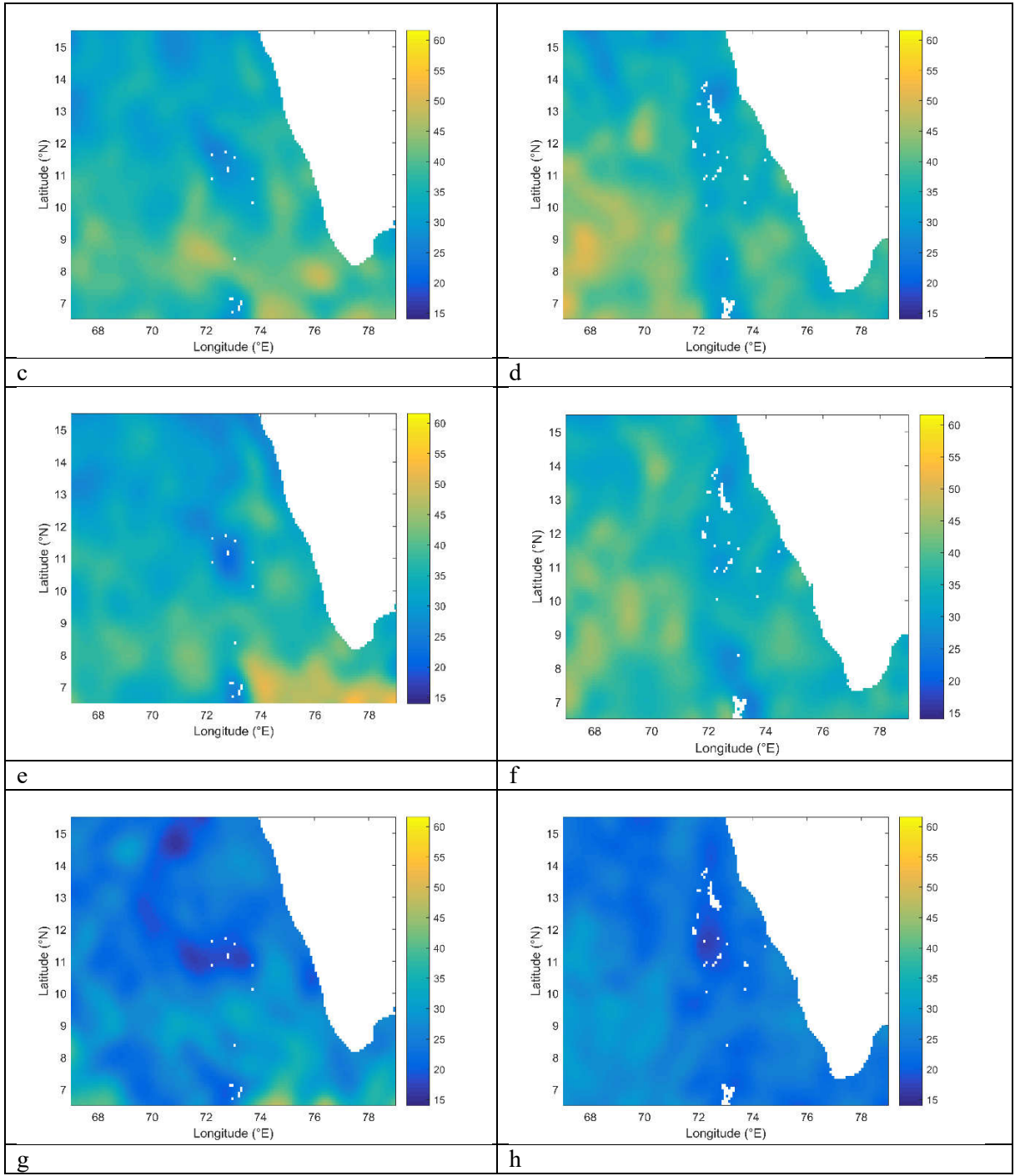

**Fig.2** Spatial distribution of the short correlation length at the sea surface (left panels) and 156 m depth (right panels) for the following variables : (a,b) temperature, (c,d) U-component of current velocity, (e,f) V-component of current velocity, (g,h) salinity

Figure 2 shows that the values of the short correlation length are similar at different depth levels and different field variables: T, S, U, V. We are interested in the coherent structures at meso and sub-mesoscale ranges which penetrate deep into the ocean interior, sometimes down to 1000 m. Therefore, to protect the consistency of calculations the same spatially varying value of correlation length calculated for the surface temperature was used for all variables and all depths of the child model grid. The validity of such simplification can be judged by model validation shown later in the text.

The downscaled value for the variables $\widetilde{Q_m}$, where $m$ is a node number on the child mesh is calculated using the equation (Gandin1965; Kalnay 2003)

$$\widetilde{Q_m} = \sum_{ni \in CI} P_{ni,m} \widetilde{Q_{ni}}, \tag{4}$$

where $CI$ is the set of nodes in the area of influence, approximately six correlation lengths in diameter and the weighting factors $P_{ni,m}$ are obtained from the solution of the system of linear equations (Gandin 1965) which is extended to take into account slow spatial variations of the correlation length

$$\sum_{ni \in CI} C(|r_i - r_k|, r_m) P_{ni,m} = C(|r_k - r_0|, r_m). \tag{5}$$

where, given the high density of parent data, the weights satisfy the normalisation condition $\sum_{ni \in CI} P_{ni,m} = 1$ with high accuracy. In these conditions Eq(4) provides the best unbiased linear estimate (BLUE) to the true value (Gandin 1965; Kalnay 2003; Gusarov et al. 2017).

The system in Eq. (5) is solved for all child model grid nodes $m$. This is the most computationally expensive part of the method as there are 764,858 nodes and approximately 118,000,000 weighting factors in the child LD20_SDD model, which is done separately for four field variables. The relieving fact is that the weighting factors are computed only once for the whole forecasting/hindcasting period. The estimate of the full value of the variable $\widetilde{Q_m}$ is obtained utilising the local homogeneity of the first statistical moments. The process is applied separately for potential temperature, salinity, northward and eastward components of velocity and for each day during the period from 01-01-2015 to 31-12-2018.

### LD20-NEMO model

The model is based on NEMO - Nucleus for European Modelling of the Ocean version 3.6 (stable) ocean modelling engine (Madec et al. 2017) and is set up in the same geographical area and with the same depth levels as LD20-SDD.

The model variables are discretised on the Arakawa C-grid. The model uses the variable volume non-linear free surface and the Total Variation Diminishing time stepping scheme. LD20 employs the Laplacian formulation for horizontal viscosity with the 3D time-varying diffusivities which are set using the Smagorinsky approach (compilation keys *key_ traldf_c3d* and *key_ traldf_smag*) with the multiplicative factor *rn_chsmag= 1.0.* For current velocities, LD20 uses a combination of Laplacian and bi-Laplacian horizontal diffusivity with multiplicative coefficients *rn_cmsmag_1 = 1.0* and *rn_cmsmag_2 = 1.0* for the Laplacian and bi-Laplacian components respectively. Vertical diffusion and viscosity coefficients are calculated using the *k-ε* option in the General Length Scale (GLS) turbulence closure scheme (*key_zdfgls*). The baroclinic and barotropic time steps are 120 and 6 seconds respectively. The model bathymetry was obtained from GEBCO_2020 Grid (GEBCO 2020) with 15 arc-second resolution. The model is forced by U and V wind speeds at 10 m above surface, air temperature at 10 m above surface, total downward shortwave radiation flux, total longwave radiation flux, precipitation and relative humidity. The wind stress and surface radiation fluxes are estimated using the CORE formula of Large and Yeager (2004). The meteorological forcing is extracted from the global atmospheric data set product NWP768 (Walters et al. 2014). The LD20 is run operationally within ROSE-CYLC model control suite (Oliver et al. 2019; ROSE 2019). The initial state and the lateral open boundary conditions are taken from the same parent model as LD20-SDD. They are implemented using NEMO unstructured BDY algorithm (Madec et al. 2017), including Flather radiation conditions for barotropic components and flow relaxation scheme (FRS) for baroclinic velocities. The width of the sponge layer for FRS is ten grid nodes. The current velocities at the boundary from the parent model are combined with tidal currents produced by nine tidal harmonics obtained from TPXO version 7.1 (Egbert and Erofeeva 2002). The first guess model tuning parameters are taken from (Bruciaferri et al. 2020) and further adjusted from the comparison of model results against the Operational Sea Surface Temperature and Ice Analysis (OSTIA) data base (OSTIA 2022). The LD20-NEMO model outputs 3-hourly instantaneous and daily average values for temperature, salinity and other oceanographic parameters.

## Results

Both LD20-SDD and LD20-NEMO models were run independently for the period from 01-01-2015 to 31-12-2018. The LD20-SDD was run on an office Windows PC, while LD20-NEMO was run of an HPC cluster using 96 computing cores.

The skills of LD20-NEMO and LD20-SDD have been assessed by comparing the SST produced by the models as well as CMEMS parent model and GHR-MUR observations against OSTIA data. The following parameters have been computed for each day of the study period: area averaged temperature for each data set ($AVG\_SST\_M$ and $\_SST\_OSTIA$), area averaged bias ($BIAS\_SST\_M$), root-mean-square deviation ($RMSD\_SST\_M$), root-mean-square deviations of anomalies ($RMSDA\_SST\_M$), and the correlation coefficient $r$ using equations (6)-(13) below

$$AVG\_SST\_M = \langle SST\_M(m) \rangle \qquad (6)$$

$$AVG\_SST\_OSTIA = \langle SST\_OSTIA(m) \rangle \qquad (7)$$

$$BIAS\_SST\_M = AVG\_SST\_M - AVG\_SST\_OSTIA \qquad (8)$$

$$RMSD\_SST\_M = \sqrt{\langle [SST\_M(m) - SST\_OSTIA(m)]^2 \rangle} \qquad (9)$$

$$RMSDA\_SST\_M = \sqrt{\langle [SSTA\_M(m) - SSTA\_OSTIA(m)]^2 \rangle} \qquad (10)$$

$$SSTA\_M(m) = SST\_M(m) - AVG\_SST\_M \qquad (11)$$

$$SSTA\_OSTIA(m) = SST\_OSTIA(m) - AVG\_SST\_OSTIA \qquad (12)$$

$$CORR = \frac{\langle SSTA\_M(m) \cdot SSTA\_OSTIA(m) \rangle}{\sqrt{\langle (SSTA\_M(m))^2 \rangle \cdot \langle (SSTA\_OSTIA(m))^2 \rangle}} \qquad (13)$$

where symbol '$\_M$' is a placeholder for the name of the data set used for comparison with OSTIA (M being one of CMEMS, LD20-NEMO, LD20-SDD, GHR-MUR), $m$ is the node number on the child model grid at the surface, and the subscripted angle brackets $\langle \rangle$ denote the area average, symbol 'A$\_$' denotes the anomaly around the area average. Area averaging takes place over the LD20 model domain but excluding a narrow flow relaxation sponge rim used by LD20-NEMO (approximately 90 km in width) around the open boundaries. The mismatch between two observational data sets, OSTIA and GHR-MUR provides a reference for assessing the quality of models.

The time series of area averaged Sea Surface Temperature from LD20-SDD, LD20-NEMO, OSTIA and GHR-MUR is shown in Figure 1.

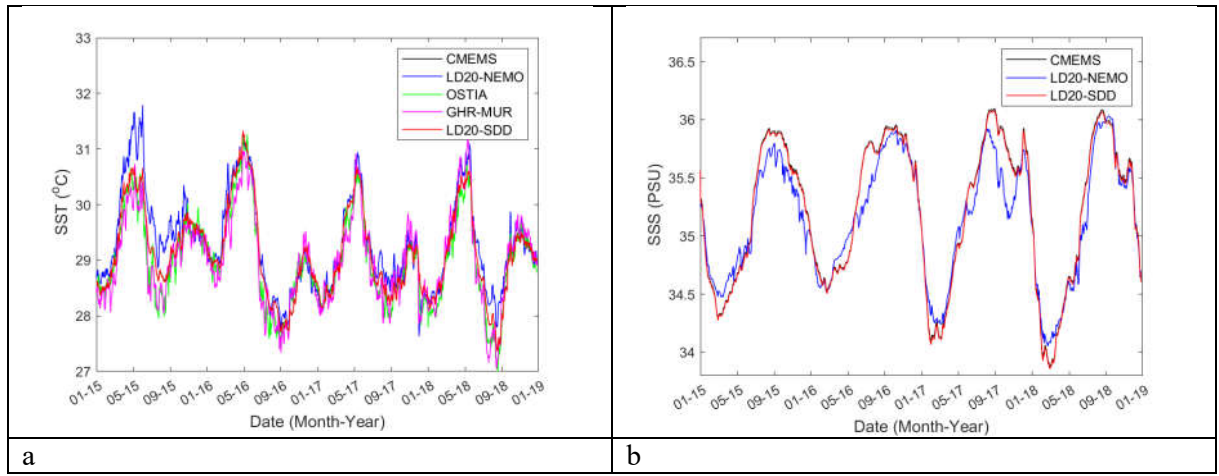

**Fig. 3** Time series of area averaged surface temperature and surface salinity for different data sets for the period 1-1-2015 to 1-1-2019. (a) Surface temperature for the numerical models CMEMS, LD20-NEMO and LD20-SDD together with data products based on measurements OSTIA and GHR-MUR, (b) Surface salinity for CMEMS, LD20-NEMO and LD20-SDD

The large seasonal variability in temperature and salinity in the Lakshadweep Sea is consistent with observations carried out during the Arabian Sea Monsoon Experiment (ARMEX), see (Gopalakrishna, et al. 2005).

The time series of model skill parameters specified by Eqs (6)-(13) is shown in Figure 4

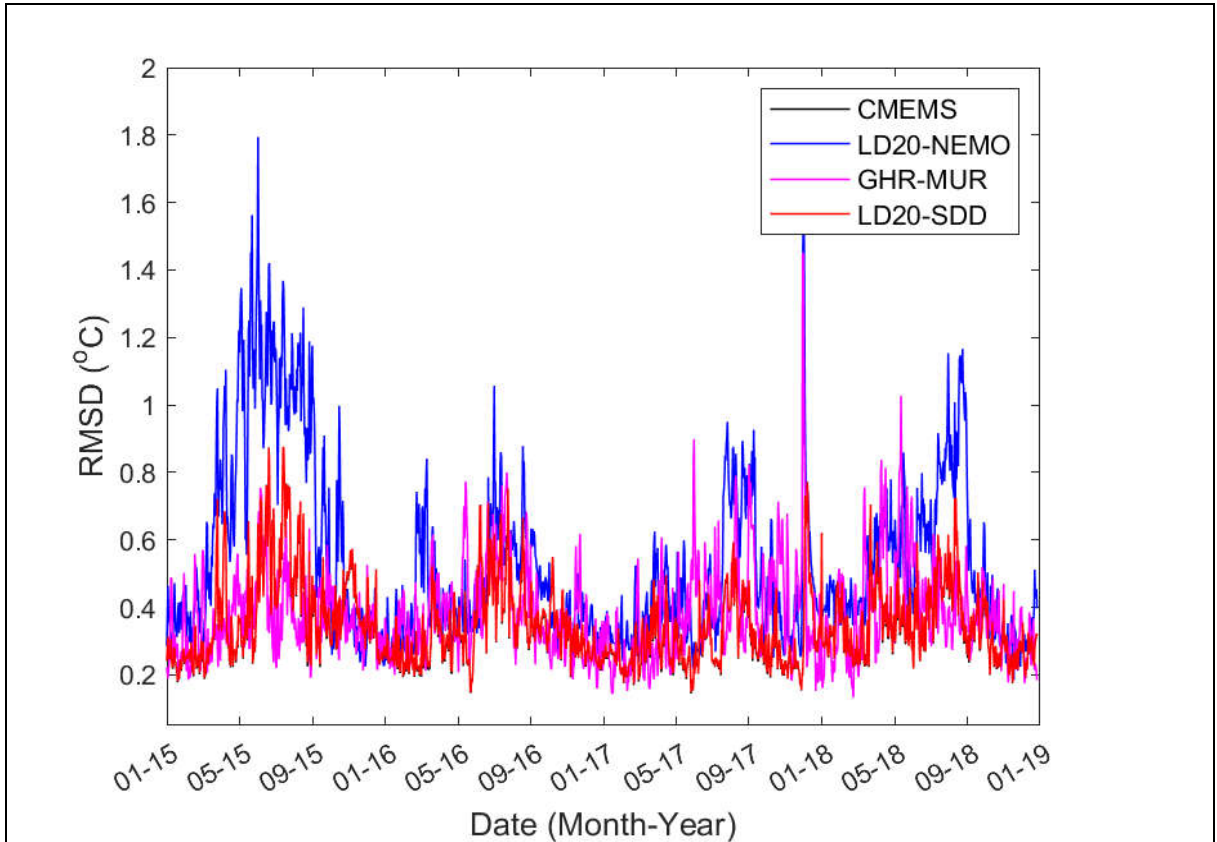
a

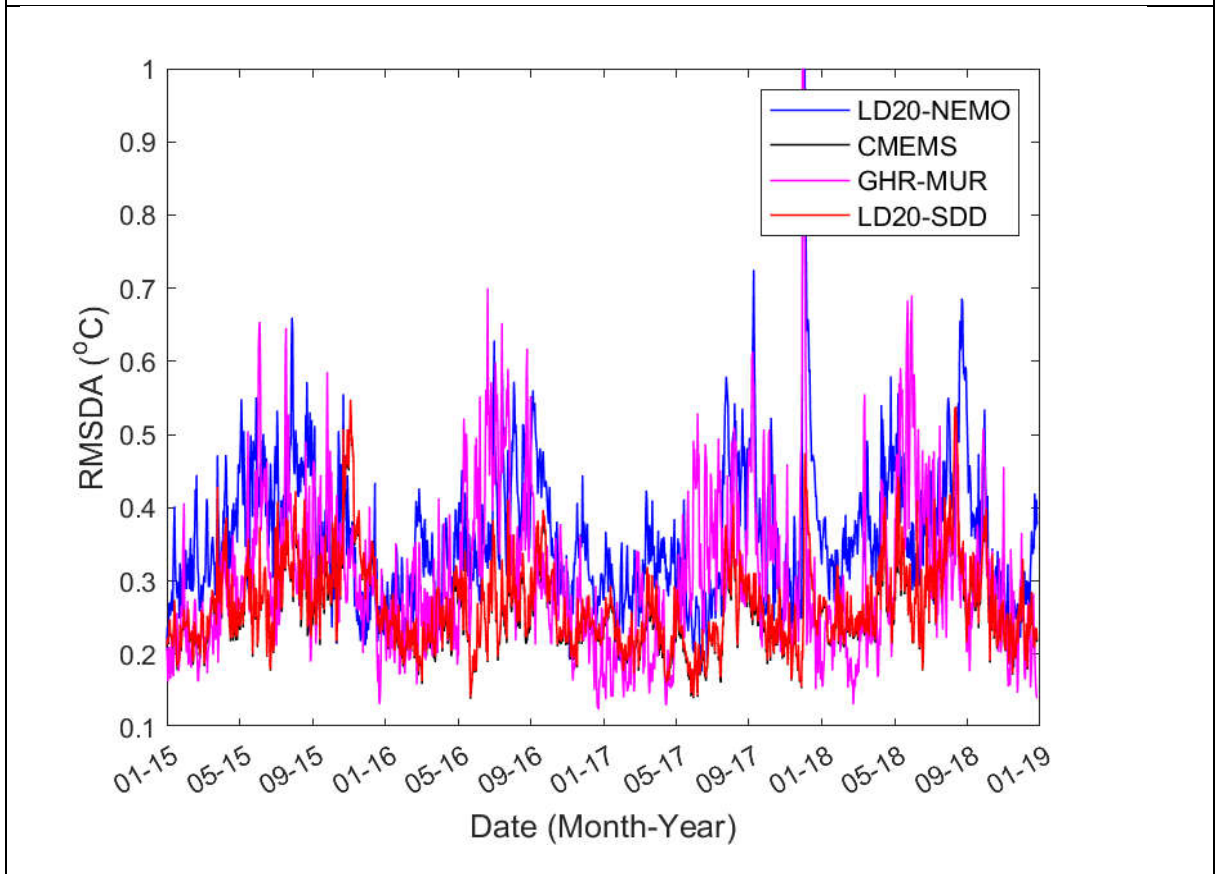
b

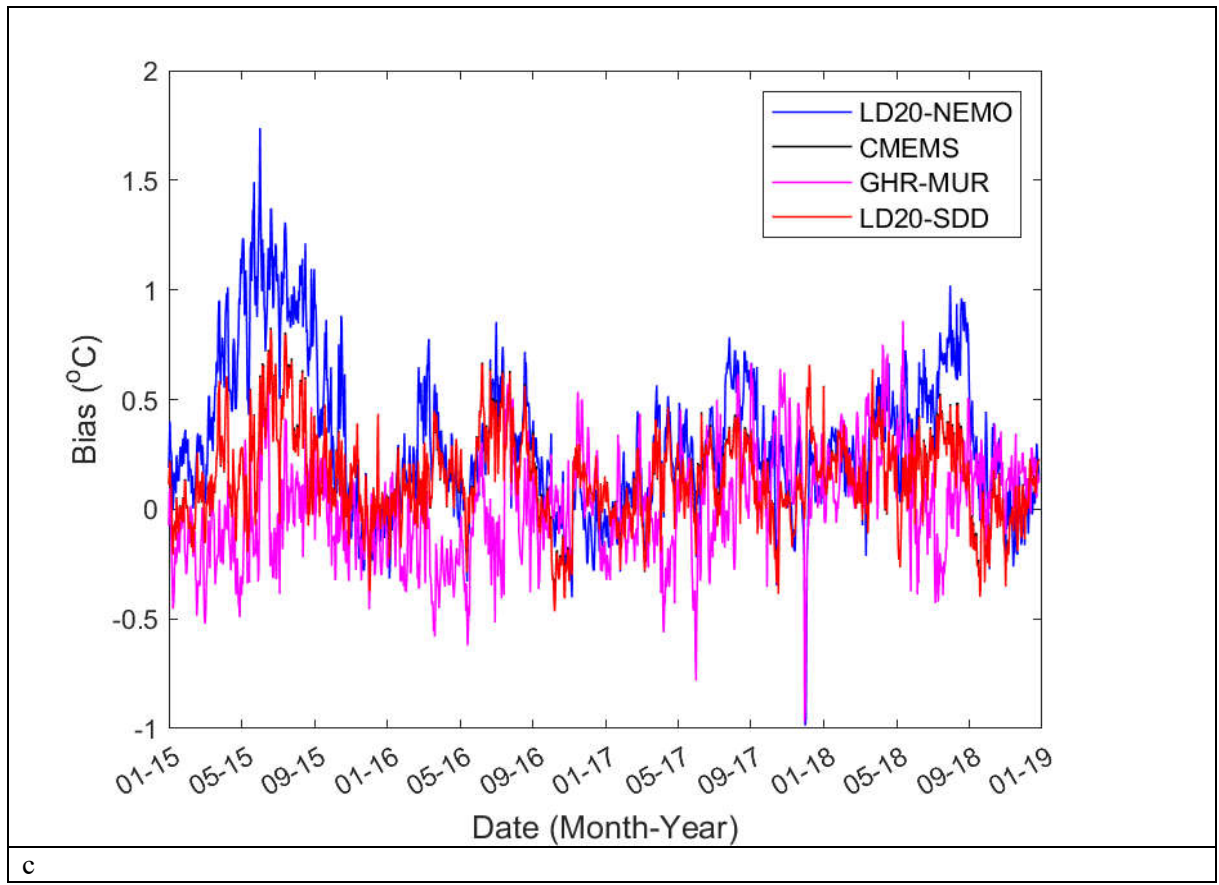

**Fig. 4** Time series of skill parameters for the temperature for different data sets (LD20-SDD, LD20-NEMO, CMEMS and GHR-MUR) compared to OSTIA: (a) RMSD for actual values, (b) RMSDA for anomalies, (c) bias. Skill parameters are defined in equations (6)-(13)

All three models, CMEMS, LD20_NEMO and LD20_SDD have approximately the same daily deviations from observations as the deviations between the two observational data sets, OSTIA and GHRSST-MUR except for the year 2015. In the warm season of this year, the LD20_NEMO produces slightly higher SST than other model and observations. The comparison of Figures 4 (a-c) suggests that the difference is caused by a positive temperature bias during this period. This suggestion is also supported by the fact that the time series of the debiased deviation represented by RMSDA for LD20_NEMO is in line with other models on observations. The effect of overestimated solar radiation in the meteorological forcing is evident for the year 2015 and it disappears after the correction to the NWP768 radiance data was made by the data provider from 15.03.2016 (MetOffice, 2017). Therefore, Table 1 below shows the year 2015 separately as well as in combination with other years.

*Table 2* **Time averaged RMSD, RMSDA, BIAS and CORR for GHR-MUR, CMEMS, LD20-NEMO and LD20-SDD with OSTIA SST taken as reference**

|  |  | GHR-MUR | CMEMS | LD20-NEMO | LD20-SDD |
|---|---|---|---|---|---|
| **Average over year 2015** | **RMSD** (°C) | 0.37 | 0.39 | 0.70 | 0.39 |
|  | **RMSDA** (°C) | 0.31 | 0.29 | 0.37 | 0.29 |
|  | **BIAS** (°C) | -0.07 | 0.18 | 0.53 | 0.17 |
|  | **CORR** | 0.59 | 0.53 | 0.44 | 0.50 |
| **Average over years 2016-2018** | **RMSD** (°C) | 0.39 | 0.33 | 0.47 | 0.33 |
|  | **RMSDA** (°C) | 0.31 | 0.26 | 0.36 | 0.26 |
|  | **BIAS** (°C) | 0.04 | 0.12 | 0.23 | 0.12 |
|  | **CORR** | 0.70 | 0.68 | 0.55 | 0.65 |
| **Average over years 2015-2018** | **RMSD** (°C) | 0.38 | 0.35 | 0.53 | 0.35 |
|  | **RMSDA** (°C) | 0.30 | 0.26 | 0.36 | 0.26 |
|  | **BIAS** (°C) | 0.01 | 0.14 | 0.31 | 0.14 |
|  | **CORR** | 0.67 | 0.65 | 0.53 | 0.61 |

In Table 3, a better performance is indicated by lower values of BIAS, RMSD, RMSDA and higher values of correlation coefficient CORR. In terms of temperature anomalies represented by RMSDA, the deterministic child model LD20-NEMO produces approximately the same level of discrepancy against OSTIA as the alternative observational data set GHR-MUR and the parent model CMEMS over all three time periods shown in Table 4. As expected, the bias in LD20-NEMO is higher in the year 2015 (0.53 °C), and then it reduces in the subsequent years to 0.23 °C, giving the four-year average of 0.31 °C. The elevated bias contributes to higher values of root-mean-square deviations in the first year (0.70 °C) which then reduces to 0.47 °C. The correlation coefficient, which is independent of the bias is still somewhat lower in LD20-NEMO than in the observational GHR-MUR and coarser model CMEMS. The slight deterioration of correlation in LD20-NEMO is probably due to the 'double-penalty' effect, which generates higher RMSD errors caused by small spatial shift in the distribution of field variables in finer-resolution models (Zingerlea and Nurmib, 2008).

The stochastic model LD20-SDD consistently shows slightly smaller (better) bias, RMSD and RMSDA errors as well as higher correlation with observations than the deterministic LD20_NEMO. The discrepancy between LD20_SDD and OSTIA are similar to, and sometimes better than the differences between observational data sets, GHR-MUR and OSTIA, except for the bias. The area and time averaged bias between any of the three models and OSTIA is higher than between the observational data sets.

The skills of models at sub-surface depth levels are assessed by comparing with ARGO float profiles. Figure 5 shows comparison of temperature and salinity profiles from CMEMS, LD20-NEMO and LD20-SDD against Argo observations. The Argo profiles cover the period from 1-1- 2015 to 31-12 2018, in total there are 325 profiles in the Lakshadweep Sea. The profiles from the models are interpolated to Argo observations in time and space, and the differences are calculated at the common depth levels taken from the Copernicus CMEMS model. The model skill parameters ( Bias, RMSD, RMSDA) are computed using averaging over four years. All models show the largest uncertainty at about 200 m depth. This is likely due to overestimating of temperature and salinity by the CMEMS model, which provides either boundary (to LD20-NEMO) of full 3D data (to LD20-SDD) to other models.

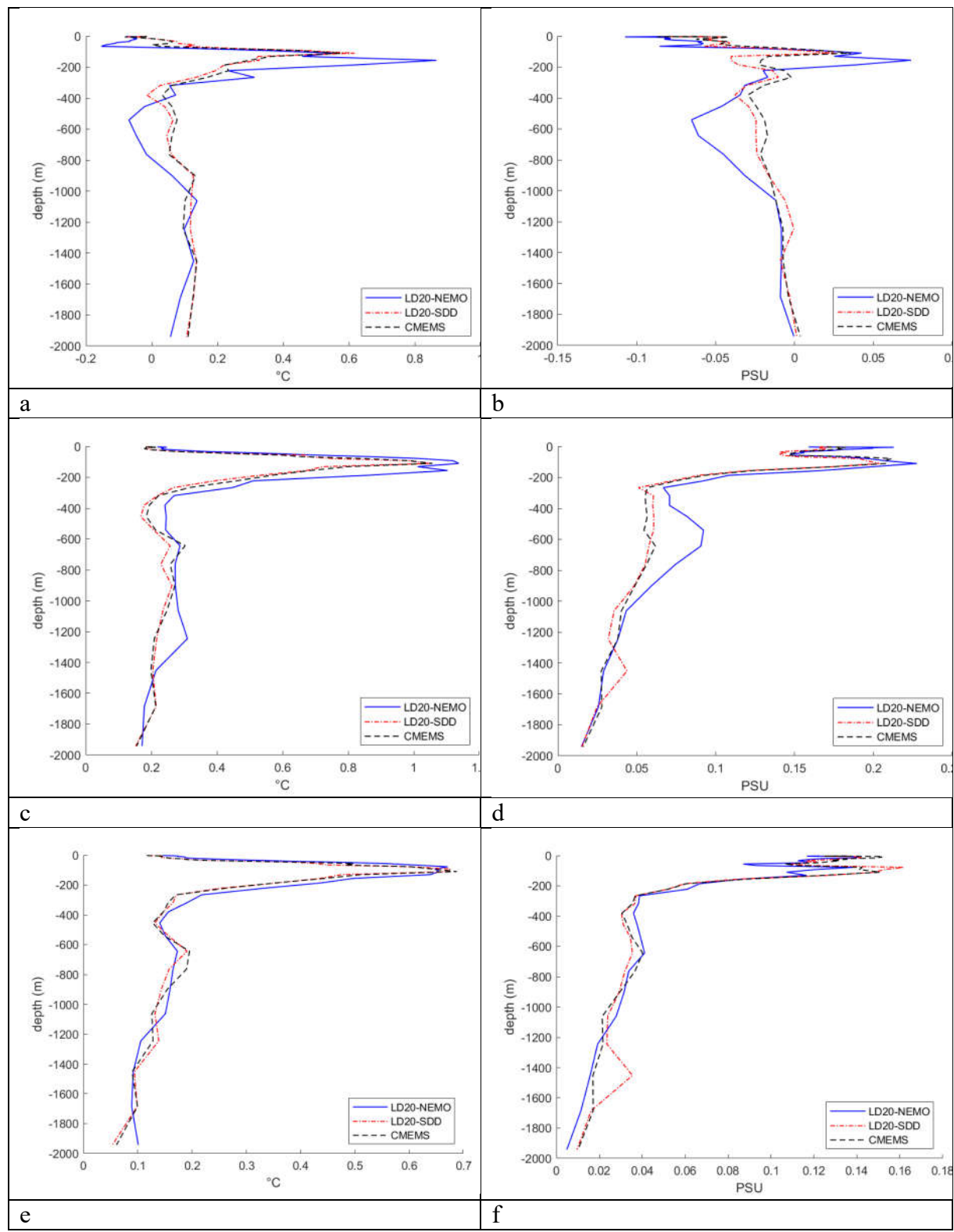

a b c d e f

**Fig. 5** Comparison of temperature (left panels) and salinity (right panels) profiles from CMEMS, LD20-NEMO and LD20-SDD models with respect to ARGO float observations:

(a,b) bias; (c,d) RMSD; (e,f) RMSDA. The period of averaging is from 01-01-2015 to 31-12-2018

The ability of a model to resolve smaller scale features can be assessed by analysing the simulated fields of relative vorticity. Vorticity is an important characteristic of the mesoscale and sub-mesoscale dynamics of the ocean and is a powerful tool to analyse ocean dynamics (Chassignet and Xu 2017). Vorticity is calculated using derivatives of current velocity, and hence an overly-smoothed representation of velocity will result in underestimation of vorticity. Figure 6 shows a snapshot of the surface velocity and vorticity fields produced by CMEMS at 1/12°, as well as LD20-NEMO and LD20-SDD at 1/20°.

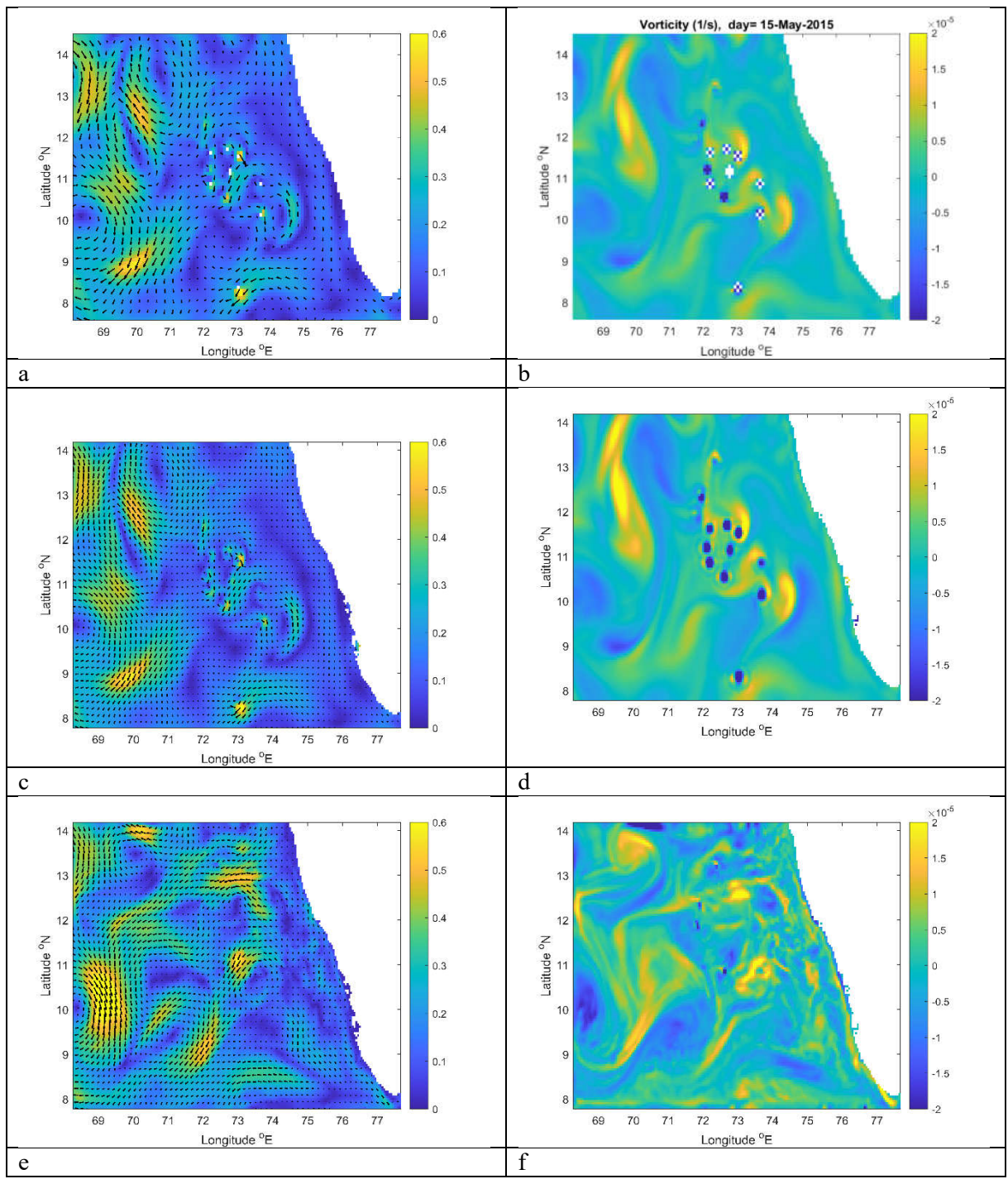

**Fig. 6** Velocity (left panels) and vorticity (right panels) computed by CMEMS (a,b), LD20-SDD (c,d) and LD20-NEMO (e,f) models on 15 May 2015.

The larger-scale velocity fields are similar between the three models, and the differences are better seen on the vorticity maps. The LD20-SDD model gives higher values of vorticity than CMEMS and resolves some smaller scale features which are only embryonically seen in CMEMS, in particular in the NW corner of the domain and around the islands. The LD20-NEMO model also gives higher values of vorticity, however the spatial pattern is more chaotic than in other models. We are not aware of any high resolution in time and space data on the velocity or vorticity in this area, therefore we can only judge indirectly the validity of patterns represented by the higher-resolution models. The CMEMS model is data assimilating and the results have come from a reliable source, therefore it is reasonable to consider the larger scale patterns from this model as reference. The vorticity patterns from LD20-SDD are more consistent with CMEMS than patterns from LD20-NEMO. The spatial shift and deformations of vorticity pattern in LD20-NEMO could be caused by the 'double penalty' effect which is common to higher-resolution models.

The ability of finer-resolution models to better represent small scale gradients is seen from the time series of area averaged enstrophy. For this analysis enstrophy (the square of vorticity) is a more suitable variable than vorticity itself. The areal integral of vorticity is simply a linear integral of normal component of velocity along the boundaries which is the same for all models as they take boundary data from the same source (CMEMS). Figure 7 shows how the area-averaged enstrophy varies with time. The enstrophy was calculated using daily data from the three models.

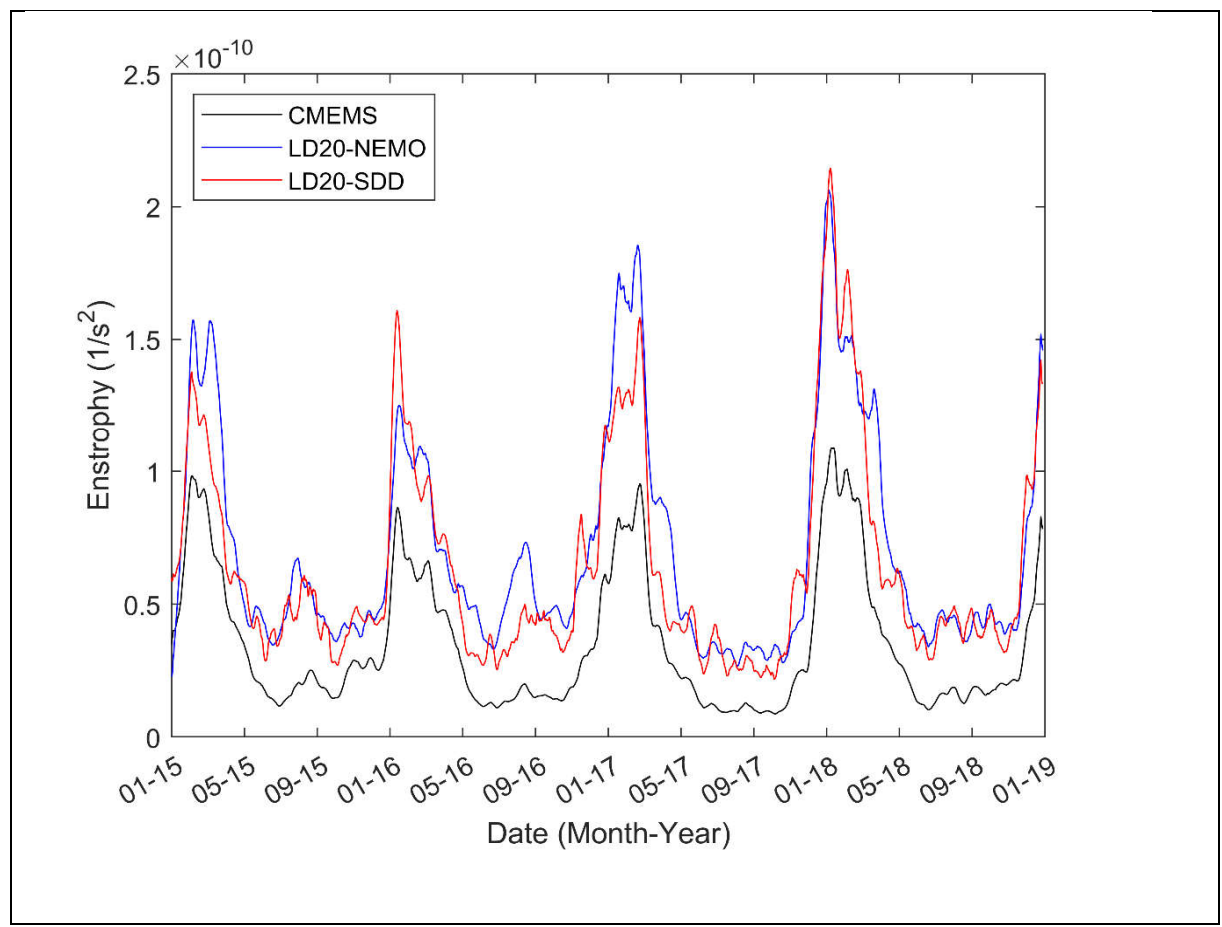

**Fig. 7** Time series of area averaged enstrophy computed by CMEMS, LD20-NEMO and LD20-SDD models for the period 1-1-2015 to 31-12-2018

An important benefit of higher resolution models is to better represent non-linear dynamics and ageostrophic flows (Chassignet and Xu 2017). The role of the non-linear effects can be assessed by the Kibel number *Ki* (Gill 1986; Phillips 1963; Vallis 2017) which is equal to the ratio of the absolute values relative to planetary vorticities. In a curved flow, such as a circular eddy, the use of geostrophic formulas leads to either underestimation (in anticyclones) or overestimation (in cyclones) of orbital velocity due to omission of the ageostrohic component of the current caused by the centripetal force (Holton and Hakim 2013). The Kibel number in this case is equal to the ratio of ageostrophic to geostrophic velocity.

In order to separate areas of high and low non-linearity in the ocean dynamics, we set a threshold value of the Kibel number Ki=0.5. Figure 8 shows the time series of Kibel number for year 2017 which reveals that the mesoscale activity exhibits strong seasonal variability being higher in winter-early spring. Similar variability was detected in other years of study.

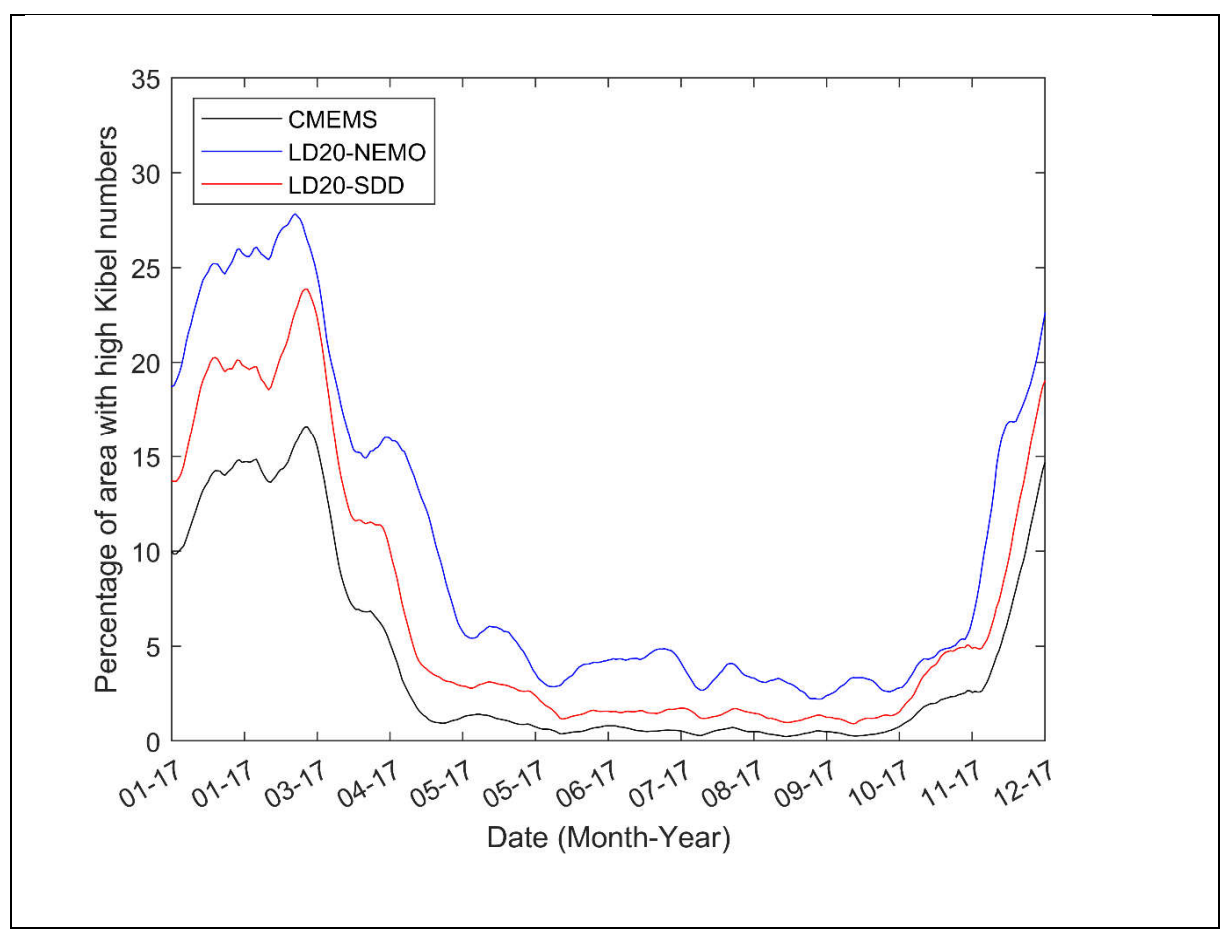

**Fig. 8** Seasonal variability of areas occupied by highly non-linear processes ( Ki>0.5) as obtained from CMEMS, LD20-NEMO and LD20-SDD models for the period 1-1-2017 to 31-12-2017

Figure 8 shows that the mesoscale activity has a strong seasonal variability being stronger in winter-early spring. Figure 9 below shows the maps of high ( Ki>0.5) and low (Ki<0.5) non-linearity in the Lakshadweep Sea in winter and summer. The highest and lowest percentage of areas occupied by non-linear dynamics is given by LD20_NEMO and CMEMS respectively. The LD20_SDD model gives intermediate values. It likely that CMEMS underestimates the size of high non-linearity areas due to insufficient resolution. In early spring, the area occupied with highly non-linear processes can be as high 20% or more of the whole Lakshadweep Sea.

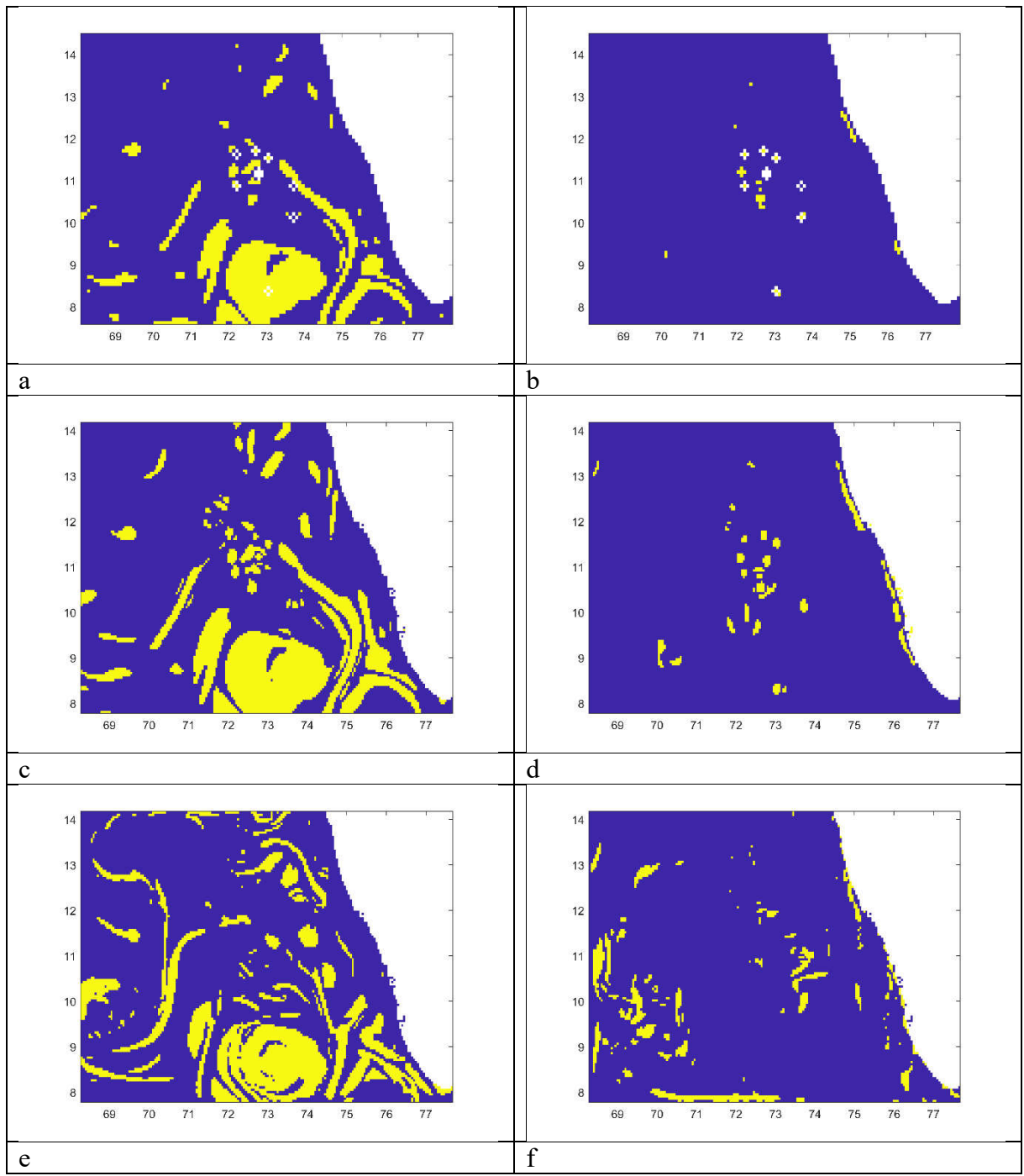

**Fig. 9** Maps of areas with high (Ki>0.5, yellow) and low (Ki<0.5, violet-blue) non-linearity in the Lakshadweep Sea in winter (left) and summer (right) for models CMEMS (a,b), LD20-SDD (c,d), LD20-NEMO (e,f)

The computational efficiency of the higher-resolution models is as follows. For LD20-NEMO, one model day of simulation takes 2.5 minutes on 96 computing cores of an HPC cluster. LD20-SDD is run on a single core of an office Windows PC. It also takes 2 min to simulate one model day, the speed is mostly dependent on the speed of reading and writing data to the disk storage. Therefore, the LD20-SDD model is approximately 100 times more computationally efficient than LD20-NEMO.

## Discussion

The need for higher resolution ocean modelling has been identified in several areas of research. The study of mesoscale and sub-mesoscale dynamics in the Gulf of Aden was much improved when 5 km and 1.5 km resolution models were used even though the baroclinic radius of deformation was in the range of 40 to 50 km (Morvan et al. 2020).

The increase of ocean resolution in global coupled models, where the ocean component explicitly represents transient mesoscale eddies and narrow boundary currents, was shown to improve the coupled ocean-atmospheric model and give a better weather forecast (Hewitt et al. 2017). The uncertainty of climate models is partially attributed to the insufficient resolution of their ocean component (Purcell 2019; Cheng et al. 2016) However, the refinement of model resolution is associated with significant increase in computational cost (Hewitt et al. 2017). In this study we provide a comparative skill test for two ocean models of the same resolution, LD20-NEMO and LD20-SDD. The former is based on the widely used deterministic approach, while the latter is based on the new Stochastic-Deterministic methodology and is significantly faster, by a factor of about 100. The models were set up in the Lakshadweep Sea which is known for its dramatic seasonal change of general circulation and intensive mesoscale dynamics in particular around the southern tip of India (Pednekar 2022). The Lakshadweep Sea is an important source of food supply for India, Sri Lanka and the Maldives (Dhaneesh et al. 2012), and the efficiency of fishery is reliant on the smaller scale phenomena such as variations in the coastal current and upwellings.

At a glance, the NEMO and SDD versions of the LD20 Lakshadweep Sea model seem to be very different. LD20-NEMO is based on the laws of physics implemented as deterministic equations whilst LD20-SDD is entirely data driven. At a deeper level, these models have some common features. Scientific laws are generalizations about a range of natural phenomena, sometimes universal and sometimes statistical (Encyclopedia Britannica 2022). Let us recall the immense catalogue of astronomical observations giving the positions of about 1,000 stars which was collected by Tycho Brahe over many years. Brahe's observations were then consolidated in the Kepler's laws of planetary motions and then further generalised by Isaac Newton as a law of gravity. However, due to the limitations of the purely deterministic approach, the equations used in ocean modelling must be supported by observations, for example in the form of data assimilation. Data assimilation uses statistical properties of innovations defined as differences between the model and observations.

In contrast to data assimilation, the SDD method is concerned with the statistics of the external data alone. The practical statistical parameter used in the SDD is the correlation length between fluctuations of field variables. Dynamical process in the ocean are interconnected and the field variables may have a number of correlations lengths reflecting different processes. The purpose of the SDD method is to reconstruct the meso- and submesocale structure of the field variables, which is only embryonically seen in the coarser parent model. A typical size of mesoscale features such as eddies or filaments is determined by the baroclinic radius of deformation, usually of the first or second mode (Robinson,2012; Koshlyakov and Monin 1978). If the non-linearity parameter of a mesoscale eddy $r = \frac{U_{orb}}{C_{phase}} > 1$ (where $U_{orb}$ is the maximum orbital velocity in the eddy, and $C_{phase}$ is the phase speed of Rossby waves) then the eddy has an inner core which traps fluid in highly correlated motion, and the outer ring which continuously entrains and sheds fluid (Klocker and Abernathey 2014). As the non-linearity parameter reduces, the inner core gets smaller. The core of an eddy gets smaller also if the eddy is embedded in a shear flow (Shapiro et al. 1997) Therefore, the area of highly correlated field variables is likely to be smaller than the area determined by the radius of deformation. This result is consistent with our calculations. In this study the correlation length is calculated using a two-scale approach (Mirouze et al. 2016). Only the shorter (i.e., mesoscale) length scale is used for further computations, as larger scale structures are well resolved by the coarser parent model. The mesoscale correlation length in the Lakshadweep Sea is typically in the range of 15 to 60 km (see Figure 2). As expected, it is lower than the first baroclinic radius of deformation in the deep Lakshadweep Sea ranging from approximately 80 to 100 km (Chelton et al. 1998).

The skills of both NEMO and SDD version of the LD20 model are assessed using the standard methodology, namely by estimating biases, two variants of RMS errors, and the correlation coefficients between the models and observation data sets. A similar comparison was performed between two alternative observational data sets: (i) GHRSST-MUR and (ii) OSTIA which gives a reference for judging the skills of LD20 models. Both LD20 models perform well, showing the deviations from the reference OSTIA data within the same range as the CMEMS reanalysis and GHRSST-MUR observational data set. However, the SDD model showed slightly better performance in terms of all skill-defining parameters.

The computations of vorticity and enstrophy show a greater difference between the lower resolution CMEMS and higher-resolution LD20 models. The under-estimation of vorticity by CMEMS is likely to be caused by the 'representative error' (Bouttier and Courtier 1999), which is related to underestimation of sharp small-scale gradients due to insufficient resolution. The 'representative error' was shown to be reduced by the SDD method which is capable of partial reconstruction of extreme values of a variable which are missed on a coarser grid (Shapiro et al. 2021).

Both SDD and NEMO versions of LD20 show higher values of enstrophy. The level of non-linearity of meso-submescale dynamics was assessed by the temporal and spatial variability of Kibel number. Higher Kibel number are associated with relatively higher ageostrophic components of current velocities. In the areas of high Kibel number the geostrophic formulas usually used to infer the surface currents from satellite derived sea level height (Mkhinini et al. 2014) may result in large errors. The areas of highly non-linear dynamics occupy as much as 20–25% of the Sea in early spring and as low as 5% in the summer. Kibel number larger than 0.5 The seasonal variability of the size of highly dynamic areas is consistent between all three models, while the LD20_NEMO gives the highest estimate (up to 28% in 2016) , LD20_SDD gives a slightly smaller figure (up to 26% in 2017) and the lower resolution CMEMS gives the lowest figures of up to 18% in 2017.

In summary, all three models give a similar representation of the area averaged values and temporal evolution of temperature and salinity. The benefit of higher resolution models come into play in simulations of gradient-dependent values, such as vorticity and enstrophy. Both versions of LD20 show higher values of vorticity and associated parameters than the coarser resolution CMEMS. The NEMO version of LD20 gives slightly higher values of enstrophy than the SDD version, however the SDD model is approximately 100 times more efficient computationally. It is difficult to judge which version of LD20 produces more realistic fields of vorticity at smaller scales as we are not aware of any current velocity observations with comparable spatial coverage and resolution.

# Conclusions

In this study we compared the skills of three ocean models, the parent model CMEMS run by EU Copernicus Marine Service at 1/12° resolution, and two child models, LD20_NEMO and LD20_SDD run at Plymouth Ocean Forecasting Centre, both at 1/20° resolution. LD20_NEMO is based on the deterministic approach while LD20_SDD uses the stochastic properties of the field variables assessed from the outputs from the parent model. All three numerical models show similar skills in reproducing temperature and salinity assessed against observations. As expected, higher resolution models better resolve smaller scale processes. This difference is particularly significant in simulation of vorticity fields and computation of the share of the sea occupied by highly non-linear processes. The SDD model is more computationally efficient than the NEMO model by a large margin.

## Funding.

This research was partly funded by the University of Plymouth through the GCRF MaMoFS project (2021)

## Data availability

The datasets generated during the current study are available from the corresponding author on reasonable request.

## **Declarations**

### Conflict of interest

The authors declare no competing interests

The Met Office Unified Model Global Atmosphere 4.0 and JULES Global Land 4.0 configurations. Geoscientific Model Development 7:361–386. doi:10.5194/gmd-7-361-2014
- Weaver AT, Mirouze I (2013) On the diffusion equation and its application to isotropic and anisotropic correlation modelling in variational assimilation. Quart. J. Roy. Meteor. Soc. 139:242–260. doi:10.1002/qj.1955
- Zingerlea C, Nurmib P (2008) Monitoring and verifying cloud forecasts originating from operational numerical models. Meteorological Applications 15(3):325–330. https://doi.org/10.1002/met.73

23